\documentclass{jaa}
\usepackage{natbib}
\usepackage{graphicx}
\usepackage{rotating, graphicx}
\usepackage{ulem}
\usepackage{hyperref}

\begin{document}\sloppy

\title{Quiescence of an Outburst of a Low-Mass Young Stellar Object: LDN1415-IRS}


\author{Koshvendra Singh\textsuperscript{1,*}, Devendra K. Ojha\textsuperscript{1}, Joe P. Ninan\textsuperscript{1}, Saurabh Sharma\textsuperscript{2}, Supriyo Ghosh\textsuperscript{1}, Arpan Ghosh\textsuperscript{2}, Bhuwan C. Bhatt\textsuperscript{3} and Devendra K. Sahu\textsuperscript{3}}

\affilOne{\textsuperscript{1}Department of Astronomy and Astrophysics, Tata Institute of Fundamental Research (TIFR), Mumbai 400005, India.\\}
\affilTwo{\textsuperscript{2}Aryabhatta Research Institute of Observational Sciences (ARIES), Manora Peak, Nainital 263001, India.\\}
\affilThree{\textsuperscript{3}Indian Institute of Astrophysics, II Block, Koramangala, Bangalore 560034, India.\\}



\twocolumn[{

\maketitle

\corres{koshvendra.singh@tifr.res.in}


\begin{abstract}
LDN1415-IRS, a low-mass young stellar object (YSO) went into an outburst between 2001 and 2006, illuminating a surrounding nebula, LDN1415-Neb. LDN1415-Neb was found to have brightened by I=3.77 mag by April 2006. The optical light curve covering $\sim$ 15.5 years, starting from October 2006 to January 2022, is presented in this study. The initial optical spectrum indicated the presence of winds in the system but the subsequent spectra have no wind indicators. The declining light curve and the absence of the P-Cygni profile in later epoch spectra indicate that the star and nebula system is retrieving back from its outburst state. Two Herbig-Haro objects (HHOs) are positioned linearly with respect to the optical brightness peak of the nebula, probably indicating the circumstellar disk being viewed edge-on. Our recent deep near-infrared (NIR) imaging using TANSPEC has revealed the presence of a nearby star-like source, south of the LDN1415-IRS, at an angular distance of $\sim$ 5.4 arcsec. \\


\end{abstract}


}]


\doinum{12.3456/s78910-011-012-3}
\artcitid{\#\#\#\#}
\volnum{000}
\year{0000}
\pgrange{1--}
\setcounter{page}{1}
\lp{1}

\section{Introduction}
Low-mass young stellar objects (YSOs) (mass $\sim$ 1-2 $M_\odot$ and age of few Myr) accrete mass in their pre-main sequence (PMS) phases. The accretion mechanism is a complex interplay among the circumstellar envelope, circumstellar disk and the central YSO. Material from circumstellar envelop falls onto the outer circumstellar disk, a case study is presented in \citet{2017ApJ...843...45K}. The loss of angular momentum of the circumstellar disk material leads to the radially inward movement of the mass \citep{1974MNRAS.168..603L,1981ARA&A..19..137P} which later follows magnetic accretion funnel \citep{1991ApJ...370L..39K,2007prpl.conf..479B} or boundary layer accretion mechanism \citep[in case of very high accretion rate,][]{1996ApJ...463..674G,1988ApJ...325..231K,2022ApJ...936..152L} to get dumped onto the star. The accretion is accompanied by mass-loss in the form of outflows/winds \citep{2003ASPC..287..333S} which can create a cavity in the nearby circumstellar cloud, giving birth to an optical nebula in case of high outflow winds. Occasionally, these YSOs are found to undergo episodes of accretion-driven outbursts followed by a long quiescent phase. \citet{10.1093/mnras/stt091} estimates a star to undergo $\sim$ 50 such outbursts during its pre-main sequence state of formation. \citet{2014prpl.conf..387A} presents an atlas of more than two dozen observed such objects. The outburst is attributed to the sudden increase in mass-accretion rate from the circumstellar disk onto the star \citep{1985ApJ...299..462H}. These objects are called FUors or EXors based on the timescales of outbursts and quiescence phases and the spectroscopic features, following the respective prototypes, FU Ori or EX Lupi. FUors show an optical outburst of 4-5 mag in a timescale of years to decades followed by a quiescence phase of decades to centuries while EXors show an optical outburst of 2-3 mag on the timescale of months to years followed by a quiescence phase of years to decades \citep{1977ApJ...217..693H,1989ESOC...33..233H}. The spectroscopic and photometric observations of these outbursting objects are crucial for understanding the mass accretion phenomena and morphology of the system as shown by  \citet{2013ApJ...778..116N,2015A&A...580A..82S} in specific cases, resolving the luminosity problem in low-mass star formation \citep{2012ApJ...747...52D}, and disk evolution where planets are to be formed \citep{2016ARA&A..54..135H} and many more.

Statistically, a very small sample of FUors and EXors is available, emphasising the careful and prolonged observation and analysis of each of such outbursting objects. In a survey of Herbig-Haro Objects (HHOs), \citet{2007A&A...463..621S} found a new optical nebula. The nebula was found in Lynds opacity class 3 cloud \citep{1962ApJS....7....1L}, LDN1415, and hence it was named LDN1415-Neb. IRAS point source catalog shows the presence of a point source, IRAS 04376+5413, well within the error ellipse (28$^{\prime \prime}$, 7$^{\prime \prime}$) of the nebula detection and hence the corresponding source of the nebula is named LDN1415-IRS. Located in Camelopardallis constellation, its coordinates are ($\alpha_{2000}$ = $04^{h}41^{m}35^{s}.94$, $\delta_{2000}$ = +54$^\circ$19$^{\prime}$16$^{\prime \prime}.87$). The distance to the nebula is 170 $\pm$ 30 pc \citep{2007A&A...463..621S}. Nebula is not detected in the optical observations prior to 2006, POSS2-red\footnote{\url{https://archive.stsci.edu/cgi-bin/dss_form}} (October 26, 1990) and KISO-I (January 22, 2001), indicating the appearance of the nebula sometime in between 2001 and 2006. The corresponding near-infrared (NIR) images from the Two Micron All Sky Survey \citep[2MASS,][]{2006AJ....131.1163S} observations (1998) showed an extended object at the location of current LDN1415-Neb. This paper outlines the results from the further follow-up post-outburst observations of LDN1415-Neb. 


In section 2, the observational details about photometry and spectroscopy are presented. Section 3 deals with the results from the data sets and discussions thereafter.
We conclude in section 4 by summarizing the important points of the paper.

\section{Observation and Data Reduction}

\subsection{Photometry}

Optical photometric observations were made with two instruments in broad-band V, R and I filters: Hanle Faint Object Spectrograph Camera (HFOSC)\footnote{\url{http://www.iiap.res.in/iao/hfosc.html}} mounted on 2-m Himalayan Chandra Telescope (HCT), IAO, Hanle, Ladakh and IUCAA Faint Object Spectrograph Camera (IFOSC) mounted on 2-m IUCAA Girawali Observatory (IGO), Pune. HFOSC provides a field of view (FoV) of 10$^{\prime}$ x 10$^{\prime}$ with a pixel scale of 0.296$^{\prime \prime}$/pixel while that of IFOSC is 10.5$^{\prime}$ x 10.5$^{\prime}$ with a pixel scale of 0.307$^{\prime \prime}$/pixel \citep{2002BASI...30..785G}. The data was collected over a span of $\sim$ 6 years starting from October 27, 2006, to March 18, 2013. In view of the declining light curve, a few recent observations were conducted with HFOSC on December 25, 26 and 30, 2021 and January 11, 2022. NIR observations included photometry in J, H and K bands. It was conducted using Near-Infrared Camera (NIRCAM)\footnote{\url{http://www.iiap.res.in/iao/nir.html}} mounted on HCT. It provided a FoV of 3.6$^{\prime}$ $\times$ 3.6$^{\prime}$ with a pixel scale of 0.4$^{\prime \prime}$/pixel. Some of the recent NIR observations included the photometry using TIFR Near Infrared Spectrometer and Imager (TIRSPEC) \citep{2014JAI.....350006N} mounted on HCT and TIFR-ARIES Near Infrared Spectrometer (TANSPEC) \citep{2022PASP..134h5002S} mounted on 3.6m Devasthal Optical Telescope (DOT). TIRSPEC provides a FoV of 5$^{\prime}$ $\times$ 5$^{\prime}$ with a pixel scale of 0.3$^{\prime \prime}$/pixel while TANSPEC provides a FoV of $\sim$ 1$^{\prime}$ $\times$ 1$^{\prime}$ with a pixel scale of 0.244$^{\prime \prime}$/pixel. The typical seeing at the IAO and IGO was $\sim$ 1.8$^{\prime \prime}$ and $\sim$ 1.2$^{\prime \prime}$ respectively during the course of our observations. Zwicky Transient Facility \citep[ZTF,][]{,2019PASP..131a8002B,2019PASP..131g8001G,2020PASP..132c8001D} archival data is also used in this study. ZTF survey uses a 48-inch Schmidt telescope at Palomar observatory. With a FoV of 47 deg$^{2}$, the facility covers more than 3750 deg$^{2}$/hour. It has three custom filters available: ZTF-g, ZTF-r and ZTF-i and has a pixel scale of 1$^{\prime \prime}$/pixel. The photometry on ZTF-r band data is done on an aperture of 10$^{\prime \prime}$ around the nebula, similar to that of other optical data. 2MASS archival data is also used. 10$^{\prime \prime}$ aperture photometry has been done on 2MASS J, H and K band images. It provided an epoch of pre-outburst NIR photometry.

The standard way of data reduction is followed by subtracting the bias frames and doing the flat-fielding with normalized flat frames in optical wavebands. In NIR, all dithered frames are median combined to get the sky background in respective wavebands which is then subtracted from the NIR source frames to remove the sky contribution followed by the flat-fielding. Aperture photometry is done for the optical nebula and its NIR counterpart in an aperture of 10$^{\prime \prime}$ in order to keep consistency with \citet{2007A&A...463..621S}. The center of the aperture is taken at the southern tip of the nebula (optical), coincident with the location of LDN1415-IRS source in NIR. For calibration, optical photometric standard stars around SA 95-41, SA 98-618, SA 98-626, SA 98-650, SA 98-961, SA 101-429, SA 113-233 and PG 0918+029 regions \citep{1992AJ....104..340L} were observed on 8 nights to obtain the atmospheric extinction and transformation coefficients. Secondary standard stars in the FoV are used to calibrate the photometric flux in the rest of the epochs. The photometric root mean square (rms) scatters for the secondary standards during the period October 2006 – March 2009 are $\leq$ 0.05, $\leq$ 0.04 and $\leq$ 0.03 in the V, R and I-bands, respectively. The magnitudes of the secondary standards matched well with the values from USNO A2.0 Catalog. NIR photometric calibrations are done by observing standard stars around AS09 region \citep{1998AJ....115.2594H} at air masses close to
that of the target observations. All reductions have been done in a standard software called Image Reduction and Analysis Facility (IRAF)\footnote{IRAF is distributed by the National Optical Astronomy Observatories, which are operated by the Association of Universities for Research in Astronomy, Inc., under cooperative agreement with the National Science Foundation.}.

\subsection{Spectroscopy}
Optical spectra were taken with HFOSC on two epochs, January 21, 2007 and November 19, 2007. The first of the two spectra had a resolution of $\sim$ 870 while the latter one has a resolution of $\sim$ 2000.
The spectra were taken by positioning the slit on the optical peak of the nebula, the southern tip of the nebula, in order to minimize the nebular contribution (see Figure 1). FeNe lamp is used for wavelength calibration. The spectra are corrected for the instrumental response using spectrophotometric standards (Hiltner 600 and G191B2B) observed on the same night. The spectra are not corrected for telluric features. All reductions have been done in IRAF. 

\section{Results and Discussion}
Using the decade-long photometric data sets and a few spectroscopic observations, the derived results are described as follows:

\subsection{Morphology}

 The nebula emerged sometime between 2001 and 2006. The aperture photometry of the nebula showed a brightening of I = 3.77 mag in comparison to the limiting magnitude of the KISO-I image taken in 2001. The nebula has an arc-shaped morphology in the south-north direction (see Figure 1). The southern part of the optical nebula coincides with a point-like object visible in the K-band images. This point-like object is LDN1415-IRS and is expected to be the driver of the LDN1415-Neb \citep[][and also see section 3.5]{2007A&A...463..621S}. 
 

The arc shape of the nebula is not uncommon in FUors. V1057 Cyg and V1515 Cyg are a few of the examples \citep{1987PASP...99..116G}. The outflow from the star, during the outburst, creates a cavity in the surrounding cloud which is illuminated by the star's light. \citet{1987PASP...99..116G} modelled these arc-shaped nebulae as an ellipsoidal illuminated cavity when viewed at an angle not equal to zero with respect to the major axis. In the model of the cavity, the star is placed at an end point of the major axis of the ellipsoid. LDN1415-IRS is also situated at the southern end of LDN1415-Neb.


\begin{figure}[!t]
\includegraphics[width=1\columnwidth]{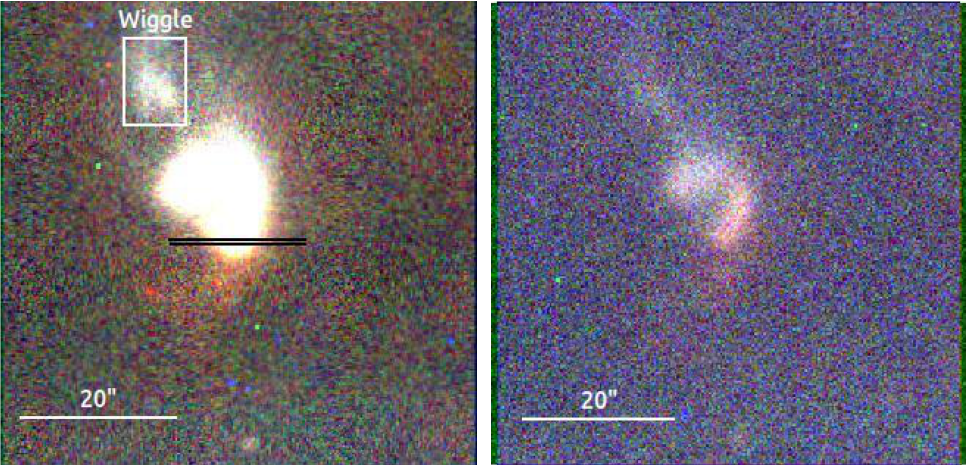}
\caption{Evolution of the optical nebula from October 28, 2006 (left) to January 11, 2022 (right) is shown by the colour-composite HFOSC images (V, R and I-band). Both of the images correspond to the same region in the sky. The two black horizontal lines in the left image indicate the position of the slit for our optical spectroscopy. North is up, and east is to left.}\label{figOne}
\end{figure}

Recent high spatial resolution (FWHM $\sim$ 1.85$^{\prime \prime}$) images with TANSPEC in J, H and K bands showed a point-like source at a projected distance of $\sim$ 5.4 arcsec south to LDN1415-IRS (see Figure 2). It will be called \textbf{LDN1415-IRS-S (S representing `southern' source)} hereafter. The nature of the second source has not yet been understood. Recent attempts for the NIR spectroscopy by HCT and DOT have failed due to the faintness of the sources. 


\begin{figure}[!t]
\includegraphics[width=0.8\columnwidth]{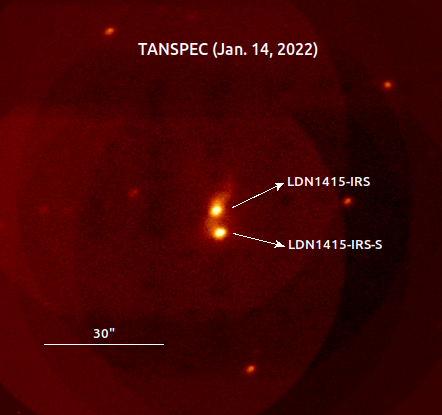}
\caption{The TANSPEC K-band image clearly shows the presence of two point-like sources. The northern one is the  LDN1415-IRS and  the southern one is the new `second source', LDN1415-IRS-S. The projected distance between the two point-like sources is $\sim$ 5.4$^{\prime \prime}$. North is up, and east is to left. }\label{figTwo}
\end{figure}

There is a `wiggly structure' extending from the northern part of the arc-shaped nebula towards the HH 892A (see Figure 1). This structure gives a glimpse of the connection between the LDN1415-IRS and the HH 892A. \citet{1989Natur.340...42R} has also argued for an HHO, HH111 to be connected to some eruptive accretion mechanism of the central star. The two HHOs, HH 892A and HH 892B,  found near the LDN1415-IRS are almost aligned in a straight line with the star from north to south (see Figure 3). The line of sight radial velocity of the two HHOs calculated from their spectra \citep{2007A&A...463..621S} was found to be similar within the error. This indicates that the star has its disk nearly perpendicular to the plane of the sky and we are facing the star-disk system nearly `edge-on'. But the similarity of the HHO's line of sight radial velocity could also be a result of the complex interaction of the HHO's with the circumstellar cloud and hence a firm bound on disk inclination angle cannot be put.

\begin{figure}[!t]
\includegraphics[width=.8\columnwidth]{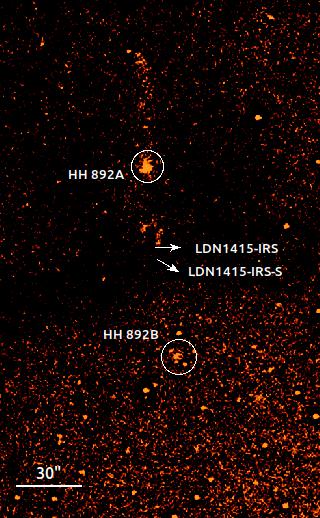}
\caption{R-band continuum subtracted H$\alpha$ image taken with HFOSC on November 19, 2007. The HHOs, HH 892A (north) and HH 892B (south) are shown in almost linear alignment with the brightness peak of the optical nebula (arcuate structure above the pointer of LDN1415-IRS). The coordinates of LDN1415-IRS and LDN1415-IRS-S are marked by the rear point of the arrows. North is up, and east is to left.}\label{figThree}
\end{figure}

\subsection{Optical and NIR variability}


Over the time of $\sim$ 15.5 years, the light curve of LDN1415-Neb has almost uniformly declined by 1.56, 2.31 and 1.72 magnitude in V, R and I bands respectively (see Figure 4). The fading of the nebula is clearly visible in the earliest and latest optical images (see Figure 1). The light curve declination in the V band (1.56) is calculated by comparing the first and the latest data from HCT/IGO while for R and I bands, it is calculated by comparing the data from \citet{2007A&A...463..621S} and the latest data from HFOSC. 

\begin{figure}[!t]
\includegraphics[width=1\columnwidth]{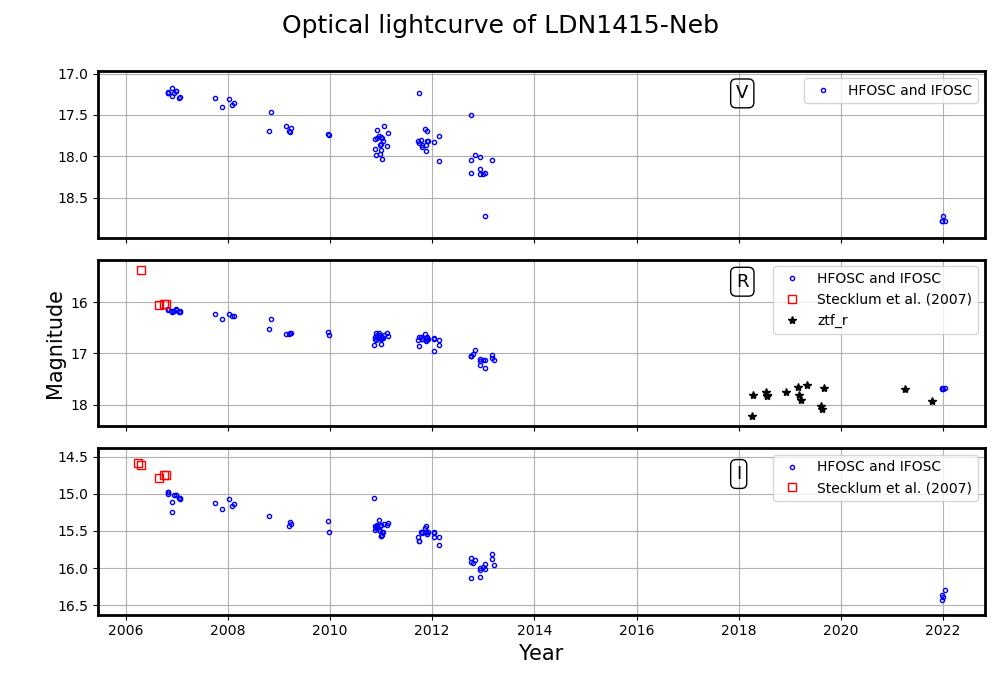}
\caption{Optical light curve of the LDN1415-Neb from April, 2006 to January, 2022. Red squares show the data from \citet{2007A&A...463..621S}. Blue circles are from HFOSC and IFOSC. Black stars represent ZTF r-band data. }\label{figFour}
\end{figure}

The LDN1415-Neb brightened by 2.45, 1.92 and 1.51 mag in J, H and K bands respectively with respect to the 2MASS NIR observations in 1998 (see Figure 5). Comparing the early epoch NIR observations on October 14, 2007 and January 11, 2022, the LDN1415-Neb has fainted by 1.01, 1.01 and 0.9 mag in J, H and K bands, respectively. The magnitude of the NIR counterpart of the 10$^{\prime \prime}$ optical nebula is in 2MASS magnitude. The J-H vs H-K color-color (CC) plot has shown the varying color of the nebular system (see Figure 6), epoch marked 10 seems to be a faulty data point. The first NIR observation after the outburst showed that the LDN1415-Neb had moved down the reddening vector from its 2MASS (1998) observations, implying the removal of A$_{V}$ $\sim$ 5 mag material from the line of sight to the LDN1415-Neb. The data points are for the NIR counterpart of the nebula, not for a single star. Since the second star is expected to be in a quiescent state (section 3.5), the color-color variation is expected to be a consequence of the first source, LDN1415-IRS. Color-Color variation can also be a consequence of variation of radiation from any other object in the system, say the disk \citep{2022ApJ...936..152L}. January 2022 observations show that it has returned to its pre-outburst stage in the CC plot. This movement of the LDN1415-Neb upward along the reddening  in the CC plot implies that the reddening has increased along the line of sight which can be interpreted as the gas/dust filling the initially created cavity. The presence of H$\alpha$ P-Cygni profile in the initial epoch spectrum \citep{2007A&A...463..621S} and absence in the later epoch spectra also support the slowing of the outflow/wind speed (see section 3.3). The slowing down of the outflow/wind can let the cavity be filled by the surrounding gas and dust, increasing the line of sight extinction and hence it could be the reason for the current position of the LDN1415-Neb on the CC plot. 

\begin{figure}[!t]
\includegraphics[width=1\columnwidth]{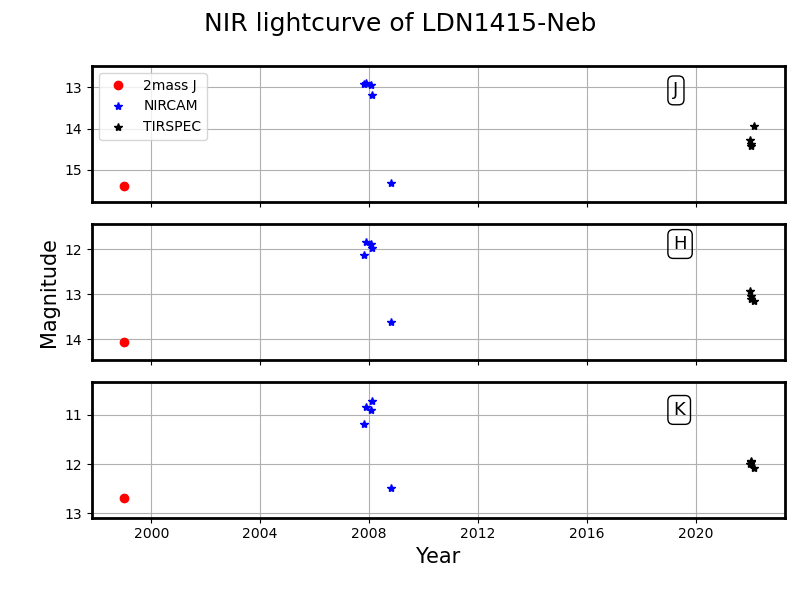}
\caption{NIR light curve of the LDN1415-Neb from October 2006 to February 2022. The red circle is from 2MASS. Blue stars are from NIRCAM. Black stars represent the TIRSPEC and TANSPEC data. }\label{figFive}
\end{figure}

\begin{figure}[!t]
\includegraphics[width=1\columnwidth]{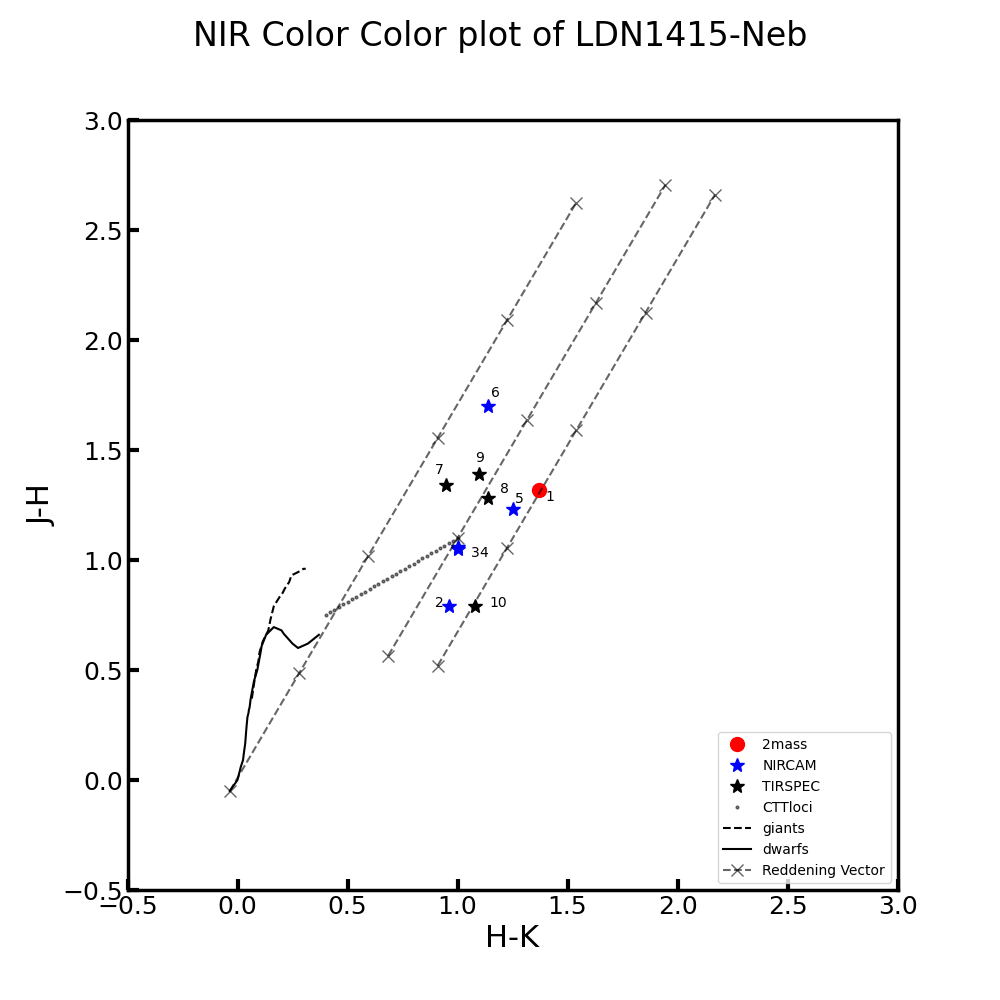}
\caption{J-H vs H-K color-color plot for the LDN1415-Neb. Red point represents 2MASS while the blue and black stars represent NIRCAM and TIRSPEC data points respectively. The sequences of field dwarfs (solid curve) and giants (thick dashed curve) are from \citet{1988PASP..100.1134B}. The dotted line represents the locus of T Tauri stars \citep{1997AJ....114..288M}. The dashed straight lines represent the reddening vectors \citep{1985ApJ...288..618R}. The crosses on the dashed lines are separated by A$_{V}$ = 5 mag. The numbers on the data points are in ascending order of their observation epochs}\label{figSix}
\end{figure}

\subsection{Optical Spectra}
One of the characterizing differences between the FUors and EXors is that the former have absorption features in the spectra but emission lines are observed in the spectra of EXors. FUors show spectra similar to that of F-G supergiants with  broadened absorption lines. Balmer H$\alpha$ P-Cygni profile is observed which is an indication of winds in the system. Broadened Li absorption at $\lambda$6707 is observed. Na I resonance line is observed at $\lambda$5890 and $\lambda$5896. Ca II infrared triplet (Ca II-IRT) is also observed with P-Cygni profiles. In the time of quiescence phases, FUors show spectra similar to that of T Tauri stars \citep{1977ApJ...217..693H,1996ARA&A..34..207H,1998apsf.book.....H}. The EXors show a plethora of emission lines such as atomic and ionic lines of Fe and He. These lines generally trace the accretion-shocked region on the star but also have the signature of the extended region (magnetospheric accretion channel). Balmer lines of H and Ca II-IRT show complex profiles, generally representing the extended region, are signatures of the accretion phenomena \citep{1950PASP...62..211H,1995A&A...300L...9L}. Outburst veils the photospheric absorption lines which become visible during the quiescence phases.

We have two epochs of the optical spectra. The recent attempts to get the optical and NIR spectra by HFOSC, TIRSPEC and TANSPEC, respectively, failed due to the faintness of the sources. The initial optical spectrum, taken by \citet{2007A&A...463..621S} (spectral resolution $\sim$ 2100), had a P-Cygni profile associated to H$\alpha$, indicating winds during the outburst, the slit covered the nebula and the HH 892A simultaneously. Our spectra taken during the later epochs have no conclusive evidence on the H$\alpha$ P-Cygni profile, the slit was placed along east-west at the southern end of the nebula (see Figure 1). No evident observation of H$\alpha$ P-Cygni profile may possibly be because of the low signal-to-noise ratio (see Figure 7) which probably indicates that the winds had slowed down (see section 3.2). Although the two spectra \citep[our and,][]{2007A&A...463..621S} were taken at two different slit positions, which might not allow for direct comparison, the non-detection of H$\alpha$ P-Cygni profile could indicate an anisotropy in the wind or slowing down of the wind by the epoch of our observation. The slowing down of the winds is supported by the decline in the light curve. In our spectra, there is no detection of CaII IRT.  [OI] line is observed at 6300 $\AA$. However, the signal-to-noise ratio is not high for this line. In Figure 8, some of the expected lines in FUors/EXors are marked by red and black vertical lines and are labelled respectively. 


\begin{figure}[!t]
\includegraphics[width=1\columnwidth]{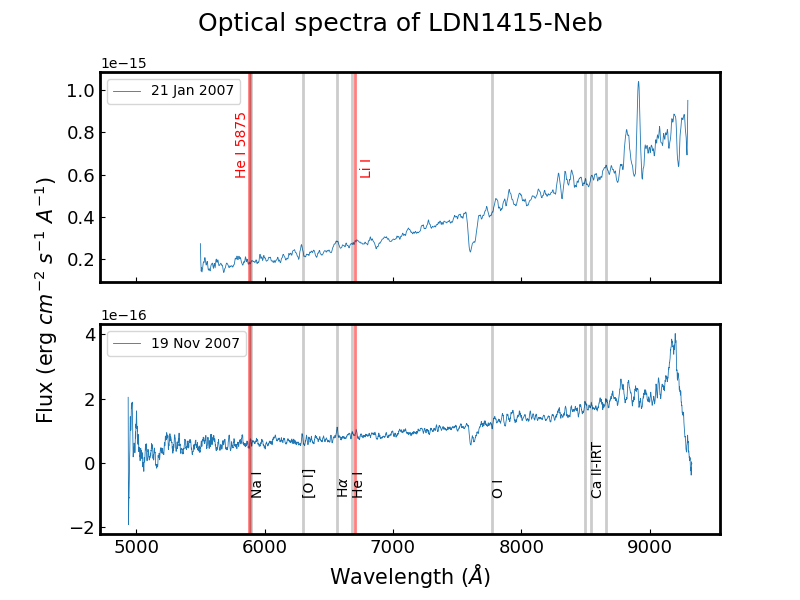}
\caption{Optical spectra of the southern tip of LDN1415-Neb taken on two epochs. Some of the expected FUors/EXors lines are marked by black and red vertical lines. Red markings are used for lines very close to black markings. Red markings are labelled by red color. The top one has a resolution of $\sim$ 870 while the bottom one has a resolution of $\sim$ 2000. The two spectra are flux-calibrated.}\label{figSeven}
\end{figure}

\subsection{Herbig-Haro Object}
The high-speed gas ejected by the stars interacts with the surrounding dust-gas cloud  and creates a shocked region of emission, called Herbig Haro Objects. HHOs emit mostly in Balmer lines and some of the low-excitation lines like [OI], [SiII], [NII] etc \citep{1983ARA&A..21..209S}.
\citet{2007A&A...463..621S} reported LDN1415-Neb while doing a survey of HHOs in Galactic dark clouds. The two HHOs, HH 892A and 892B are located in the north and south of the LDN1415-Neb respectively. Located in opacity class 3 Lynds cloud, LDN1415, the HH 892A was found to have a tangential speed of 25 kms$^{-1}$ \citep{2007A&A...463..621S} and it was moving away from the central source, LDN1415-IRS. In our later epoch H$\alpha$ images (see Figure 3), the HHOs, HH 892A (north) and HH 892B (south) are clearly visible. The presence of HHOs has given an indication of the axis-inclination but no firm bound can be put (section 3.1). They are also an identification factor for some of the violent processes like that of FUors \citep{1983ARA&A..21..209S}.

\subsection{Second Embedded Source}
2MASS catalog provides photometry for two point-like sources around the LDN1415-Neb. Interestingly, the quality flags of the 2MASS photometry of LDN1415-IRS-S are better (quality flags are UAA for JHK bands) than that of LDN1415-IRS (quality flags are UUA for JHK bands). The coordinate of LDN1415-IRS-S from the 2MASS catalog is ($\alpha_{2000}$ = $04^{h}41^{m}35^{s}.87$, $\delta_{2000}$ = +54$^\circ$19$^{\prime}$11$^{\prime \prime}.68$). A recent TIRSPEC (December 2021) image when subtracted from the first epoch NIRCAM image (2007), after convolving the two images to the same angular resolution, leaves a net detectable flux at the location of the LDN1415-IRS. This indicates that the source LDN1415-IRS is the one which probably went into an outburst. For a likely distance of 170 pc \citep{2007A&A...463..621S}, the second source is located at a minimum distance of $\sim$ 920 AU south to LDN1415-IRS. Many of the bonafide FUors/EXors are found to have a companion \citep[][and reference therein]{2014prpl.conf..387A}. To name the bonafide FUors and EXors with confirmed companions: RNO 1B/1C (FUor-like, projected linear separation = 3400 AU), FU Ori (FUor, 225 AU), AR 6A/6B (FUor-like, 2240 AU), Z Cma(FUor, 93-110 AU), V512 Per (EXor, 90 AU), UZ Tau E (EXor, 560 AU), VY Tau (EXor, 92.4 AU), V1118 Ori (EXor, 74.5 AU) \citep{2014prpl.conf..387A}. About 30$\%$ of the bonafide FUors/EXors have confirmed companions. The projected linear distance between the LDN1415-IRS and LDN1415-IRS-S (920 AU) lies in between the values reported for the rest of the bonafide FUors/EXors with confirmed companions. Companion has been proposed as a triggering agent for the accretion outburst in FUors/EXors \citep{1992ApJ...401L..31B}. According to this hypothesis, the binary star perturbs the inner disk of the first star at the perihelion in the orbit. This perturbation then leads to an outburst. This makes the LDN1415-IRS system a good candidate to be studied for any binarity and the consequences of the binarity in the outburst. 

The point spread function (PSF) photometry on the TANSPEC NIR images returns the K-band magnitude of the LDN1415-IRS and LDN1415-IRS-S to be 13.52 $\pm$ 0.03 and 13.85 $\pm$ 0.05 respectively. There are no standard stars in the FoV of the TANSPEC frame for calibration. So, the 2MASS magnitude of a star from the TANSPEC frame was calculated based on the near-simultaneous observation with the TIRSPEC at HCT, which was later used to do the zeroth order photometric correction on the instrumental magnitude (from TANSPEC) of the LDN1415-IRS and LDN1415-IRS-S. Due to the faintness of the sources, a more sensitive telescope will be essential to do the NIR spectroscopy of the two sources. With photometry and spectroscopy, a better constraint on the mass of the two stars can be put in. A constraint on the radial distance to the second star can also be put in. If LDN1415-IRS-S is established to be a companion of the LDN1415-IRS, the system can be modelled to see if a triggering of inner disk instability in the outbursting source is possible by the binary source, LDN1415-IRS-S.
Further TANSPEC photometric follow-up observations will be done for understanding the nature of LDN1415-IRS-S. With an assumption that LDN1415-IRS-S has not been active/variable, its NIR point spread function (PSF) photometry can be subtracted to remove the blending from the 10$^{\prime \prime}$ aperture photometry (see section 2.1). The disintegrated photometry of the two sources will put a better constraint on the age and mass of LDN1415-IRS. This will also enable us to put a constraint on the mass and age of the LDN1415-IRS-S. 

\section{Conclusions}
\begin{itemize}
    \item LDN1415-IRS went into an outburst sometime between 2001 and 2006 and it has been optically dimming thereafter. NIR light curve also shows a corresponding evolution to the quiescent phase.
    \item During the outburst, the outflow/wind removed material of A$_{V}$ = 5 mag from the line of sight. In recent observations, the star seems to have reddened by the same amount of material. However, heating up and then cooling down some sources within the system can also explain the color variation of the nebula during the outburst.
    \item Diminishing of the H$\alpha$ P-Cygni profile after its intense appearance indicates the decreasing strength of the outflow. This can cause the surrounding material to infall into the cavity and can cause reddening.
    \item Presence of HHOs and a nebula along with an outburst indicates that the star is a young eruptive accretion variable. The lower bound of optical outburst, I=3.77 mag and a quiescent phase of more than one and half decades indicate the star as a FUor but in absence of NIR spectra, the classification cannot be conclusively stated.
    \item Further observations and characterization of the second source we resolved in our deep NIR images are crucial to understanding the outburst trigger mechanism in LDN1415-IRS.
\end{itemize}

\section*{Acknowledgements}
We thank the staff of the HCT, Hanle and CREST, Hosakote for their support during the HFOSC, NIRCAM and TIRSPEC observations. We also thank the staff of IGO, Pune and IUCCA for their help and support during the IFOSC observations. We would also like to thank the staff of DOT, Devasthal and ARIES for their support during TANSPEC observations. 
K.S., D.K.O. and J.P.N. acknowledge the support of the Department of Atomic Energy, Government of India, under Project Identification No. RTI 4002.

\vspace{-1em}


\bibliography{reference}
\vspace{-1.5em}

\end{document}